\documentclass{article}
\usepackage{graphicx} 
\usepackage{natbib} 
\usepackage{amsmath, amssymb}
\usepackage{float}
\usepackage{subcaption} 
\usepackage{hyperref}

\defcitealias{Kingma2013AutoEncodingVB}{Kingma et al., 2013}
\defcitealias{Lucas1981AnII}{Lucas et al., 1981}

\title{Predicting risk of cardiovascular disease using retinal optical coherence tomography imaging} 
\date{}

\begin{document}

\maketitle

\author{
    Cynthia Maldonado-Garcia$^{1,*}$, 
    Rodrigo Bonazzola$^{1}$, 
    Enzo Ferrante$^{2}$, 
    Thomas H. Julian$^{3,4}$, 
    Panagiotis I. Sergouniotis$^{3,4,5,6}$, 
    Nishant Ravikumar$^{1,\dagger}$, 
    Alejandro F. Frangi$^{7,8,9,10,\dagger}$
}

\date{
    $^{1}$Centre for Computational Imaging and Simulation Technologies in Biomedicine, School of Computing, University of Leeds, Leeds, UK \\
    $^{2}$Research Institute for Signals, Systems and Computational Intelligence, sinc(i), CONICET-UNL, Santa Fe, Argentina \\
    $^{3}$Division of Evolution, Infection and Genomics, School of Biological Sciences, Faculty of Biology, Medicine, and Health, University of Manchester, Manchester, UK \\
    $^{4}$Manchester Royal Eye Hospital, Manchester University NHS Foundation Trust, Manchester, UK \\
    $^{5}$Manchester Centre for Genomic Medicine, Saint Mary’s Hospital, Manchester University NHS Foundation Trust, Manchester, UK \\
    $^{6}$European Molecular Biology Laboratory, European Bioinformatics Institute (EMBL-EBI), Wellcome Genome Campus, Cambridge, UK \\
    $^{7}$Division of Informatics, Imaging, and Data Sciences, School of Health Sciences, Faculty of Biology, Medicine, and Health, University of Manchester, Manchester, UK \\
    $^{8}$School of Computer Science, Faculty of Science and Engineering, University of Manchester, Kilburn Building, Manchester, UK \\
    $^{9}$Christabel Pankhurst Institute, University of Manchester, Manchester, UK \\
    $^{10}$NIHR Manchester Biomedical Research Centre, Manchester Academic Health Science Centre, Manchester, UK \\
    $^{\dagger}$Indicates joint senior authors.
}

\begin{abstract}
Cardiovascular diseases (CVD) are the leading cause of death globally. Non-invasive, cost-effective imaging techniques play a crucial role in early detection and prevention of CVD. Optical coherence tomography (OCT) has gained recognition as a potential tool for early CVD risk prediction, though its use remains underexplored. In this study, we investigated the potential of OCT as an additional imaging technique to predict future CVD events. We analysed retinal OCT data from the UK Biobank. The dataset included 612 patients who suffered a myocardial infarction (MI) or stroke within five years of imaging and 2,234 controls without CVD (total: 2,846 participants). A self-supervised deep learning approach based on Variational Autoencoders (VAE) was used to extract low-dimensional latent representations from high-dimensional 3D OCT images, capturing distinct features of retinal layers. These latent features, along with clinical data, were used to train a Random Forest (RF) classifier to differentiate between patients at risk of future CVD events (MI or stroke) and healthy controls. Our model achieved an AUC of 0.75, sensitivity of 0.70, specificity of 0.70, and accuracy of 0.70, outperforming the QRISK3 score (the third version of the QRISK cardiovascular disease risk prediction algorithm;  AUC = 0.60, sensitivity = 0.60, specificity = 0.55, accuracy = 0.55). The choroidal layer in OCT images was identified as a key predictor of future CVD events, revealed through a novel model explainability approach. This study demonstrates that retinal OCT imaging is a cost-effective, non-invasive alternative for predicting CVD risk, offering potential for widespread application in optometry practices and hospitals.

\end{abstract}

\section{Introduction}
Cardiovascular diseases (CVDs) continue to pose a significant global health challenge, affecting more than 500 million individuals worldwide. Specifically, in 2021, CVDs led to 20.5 million deaths. It is concerning to observe that up to 80\% of premature myocardial infarction (MI) and stroke cases could potentially be prevented if detected early. Moreover, the burden of CVDs disproportionately impacts low- and middle-income countries, where nearly four out of every five CVDs-related deaths worldwide occur \citep{Lindstrom2023}. Currently, tools like the
third version of the QRISK cardiovascular disease risk prediction algorithm (QRISK3) are utilized in primary care settings by healthcare professionals to pinpoint patients at higher risk of various CVDs. These tools are commonly employed during health assessments to assess a patient's risk based on factors such as demographic details (e.g., ethnicity, age, sex), clinical indicators (e.g., cardiac volume measurements, blood markers, indicators of obesity, etc.), and socioeconomic data \citep{Li2019RP}. Notably, early identification of individuals at risk is crucial since premature CVDs is highly preventable. Effective primary prevention strategies can lead to a decrease in CVDs mortality and morbidity, as demonstrated in several previous clinical studies \citep{Samsa1997KnowledgeOR, Kirshner2005LongtermTT, Donkor2018StrokeIT}. 

The retinal and choroidal microvasculature have been shown to be sensitive indicators of systemic vascular conditions, such as diseases affecting the cerebral and coronary vasculature \citep{Farrah2020TheET}. Therefore, retinal imaging offers a non-invasive means to scrutinize the microvasculature at the back of the eye, presenting an avenue for identifying individuals at risk of systemic vascular conditions, including stroke and heart attack. This approach enables the detection of microvascular dysfunction in peripheral vasculature, aiding in the early recognition of cardiovascular disease risk  \citep{Anderson1995CloseRO, Farrah2020TheET, Rudnicka2022ArtificialIR, DiazPinto2022PredictingMI}. Retinal imaging, being non-invasive and cost-effective, is increasingly recognized as a valuable tool for timely preventive care and effective treatment strategies \citep{Wagner2020InsightsIS}. Widely accessible in eye clinics and optometric practices, retinal imaging techniques such as fundus photography and optical coherence tomography (OCT) play a pivotal role in assessing cardiovascular disease risk  \citep{Flammer2013TheEA, Hanssen2022RetinalVD, Wong2022ArtificialII}. Ongoing research endeavors aim to further elucidate these connections and develop predictive models for early CVDs identification, propelled by advancements in retinal imaging technologies.

Numerous prior studies have investigated the application of artificial intelligence (AI) in predicting CVDs risk factors using fundus photographs \citep{Poplin2018PredictionOC, Nusinovici2022RetinalPD}, as well as CVDs events \citep{DiazPinto2022PredictingMI, Cheung2020ADS}. While retinal fundus photographs offer a two-dimensional depiction of retinal vasculature, OCT, with its high-resolution 3D imaging capabilities, allows for quantitative assessment of retinal thickness and structure, including microvasculature. This provides insights not achievable with fundus photography alone. The potential of 3D OCT imaging lies in its ability to detect subtle abnormalities in retinal microstructure and microvasculature that may go unnoticed in 2D images, thus serving as a valuable tool for identifying early disease indicators \citep{Farrah2020TheET}. This technological advancement in OCT has revolutionized retinal imaging by facilitating visualization of the chorioretinal microcirculation, which can serve as an early sign of microvascular disease. However, there remains limited research utilizing OCT as a predictor of CVDs \citep{Garca2022PredictingMI, Zhou2023AFM}.

\noindent \textbf{Contributions:} This study introduces a predictive model that integrates features derived from 3D OCT imaging through a self-supervised deep neural network, along with patient demographic and clinical details. The aim is to detect individuals at risk of MI or stroke within five years after image capture. To the best of our knowledge, this is the pioneering investigation into the application of 3D OCT imaging and artificial intelligence for automatically forecasting patients vulnerable to adverse CVD incidents. The main contributions of this research are: (i) the development of a self-supervised feature selection variational autoencoder (VAE) integrated with a multimodal Random Forest classification model, which effectively combines diverse patient data, including OCT imaging, and clinical variables; (ii) the introduction of an innovative method for enhancing model interpretability through optical flow, enabling detailed localization of retinal features that significantly contribute to the accurate identification of patients at risk of adverse cardiovascular events; and (iii) the identification of the choroidal layer as the key feature influencing the model's predictive accuracy for cardiovascular disease risk.

\section{Methodology}
\subsection{Database}

\begin{figure}[H]
\begin{center}
\includegraphics[width=\textwidth]{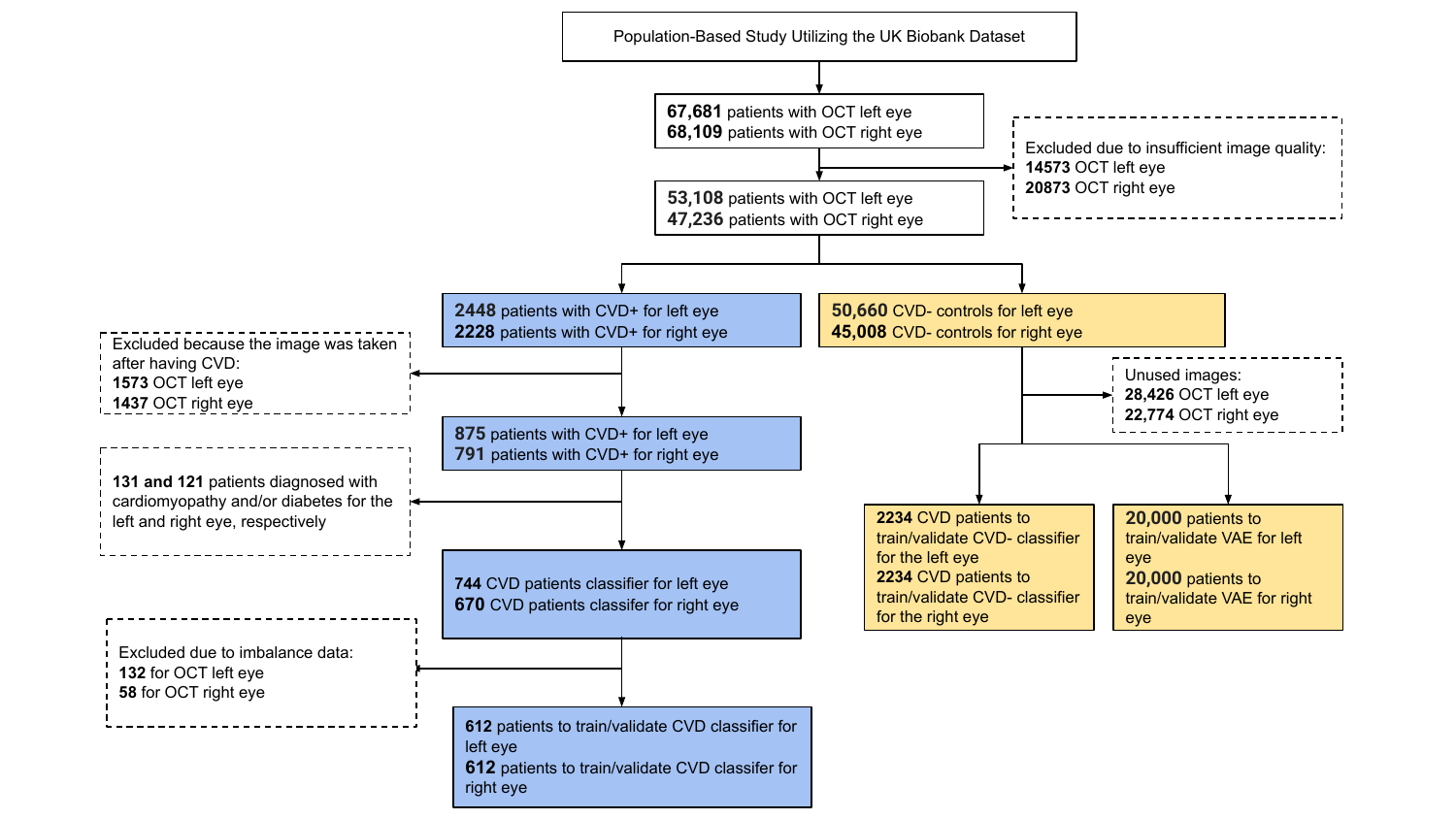}
\end{center}
\caption{ STROBE flow chart describing participant inclusion and exclusion criteria applied to define the study cohort.}\label{fig:flow_chart}
\end{figure}

In this research, we utilized retinal OCT imaging data sourced from the UK Biobank, captured using the Topcon 3D OCT 1000 Mark 2 system.  OCT, a non-invasive imaging technique, employs light waves to generate intricate images of ocular structures like the retina, choroid, and optic nerve. The imaging procedure occurred in a dimly lit environment without pupil dilation, utilizing the 3D 6×6 $mm^{2}$ macular volume scan mode, which consists of 128 horizontal B-scans arranged in a 6×6 mm raster pattern. The UK Biobank contains a vast array of health-related information from more than 500,000 participants in the UK, encompassing genetics, demographics, clinical measurements, lifestyle aspects, and medical imaging. Specifically, during their baseline visit (Instance 0, "Initial assessment visit (2006-2010)"), a total of 68,109 and 67,681 participants underwent retinal imaging for their right and left eyes, respectively. To ensure only high-quality images were included, we automatically evaluated image quality using a quality index (QI) detailed in a prior study \citep{Stein2006ANQ}. This QI is a globally accurate quality assessment algorithm derived from the intensity ratio, which is based on a histogram covering the entire image, and the tissue signal ratio, indicating the ratio of highly reflective pixels to less reflective ones. We excluded images representing the lowest 20\% as indicated by the QI indicator, resulting in the exclusion of 14,573 images for the left eye and 20,873 images for the right eye, leaving 53,108 and 47,236 remaining images, respectively. Among these, we identified 2,448 (left eye) and 2,228 (right eye) images from participants who had experienced a stroke or MI event, referred to as \texttt{CVD+} participants. However, only images from the left eye of 875 participants and the right eye of 791 participants were taken before the CVD event. Furthermore, we omitted 131 patients with diabetes and/or cardiomyopathy for the left eye and 121 patients for the right eye. We also removed cases to address data imbalance, resulting in a final cohort of 612 subjects for both eyes. A visual representation of the participant selection and exclusion criteria used to establish the cohort for this study is depicted as a STROBE diagram in Figure \ref{fig:flow_chart}. This study exclusively encompasses patients who experienced MI or stroke within a five-year period after OCT image acquisition. The acquisition details of reported CVD incidents' sources are elaborated in Figure \ref{fig:source_cvd}.

The size of the \texttt{CVD+} group was determined based on the application of specific inclusion and exclusion criteria, as outlined in the STROBE diagram in Figure \ref{fig:flow_chart}, resulting in a final count of 612 participants with OCT images of both eyes. For the non-CVD or \texttt{CVD-} group, 2234 participants were randomly selected for OCT images of both eyes. The essential patient characteristics used to match the \texttt{CVD+} and \texttt{CVD-} groups included demographic factors and clinical measurements, which are detailed in Table \ref{t:data_caracteristics}. The average age of individuals with and without CVDs was 60.78 ± 6.47 years, showing no significant difference between the two groups. The majority of participants in the UK Biobank cohort were of white ethnicity, with similar proportions in both groups. The average body mass index (BMI) was 28.31 ± 4.45 kg/m² for those with CVDs and 27.43 ± 4.33 kg/m² for those without. In terms of blood pressure readings, individuals with CVDs had a systolic blood pressure (SBP) of 147.26 ± 19.57 mm Hg, while those without CVDs had an SBP of 145.1 ± 18.75 mm Hg. The diastolic blood pressure (DBP) was 84.75 ± 10.23 mm Hg for individuals with CVDs and 83.22 ± 9.73 mm Hg for those without. The mean level of haemoglobin A1c (HbA1c) was 36.52 ± 4.32 mmol/L for individuals with CVDs and 36.59 ± 6.61 for those without. A notable percentage of participants reported being current alcohol consumers, accounting for 90.69\% of individuals with CVDs and 91.83\% of those without. These participant characteristics for both groups are summarized in Table \ref{t:data_caracteristics}.

\begin{table}[H]
  \centering
  \renewcommand{\arraystretch}{1.8} 
  \begin{tabular}{p{4cm}p{5cm}p{5cm}}
    \hline
    Characteristics & \texttt{CVD+} & \texttt{CVD-} \\
    \hline
    Number of subjects & 612 & 2234 \\
    Age: Mean (s.d), years & 60.78 (6.47) &  60.78 (6.47)\\
    Sex: F, M \% & 29.74, 70.26 &   29.74, 70.26 \\
    Ethnicity, \% & \parbox[t]{5cm}{
      90.18 White, 4.26 Mixed, 3.93 Asian or Asian British, 0.33 Black or Black British, 0.16 Chinese, 1.15 Other ethnic group} & 
      \parbox[t]{5cm}{      
      89.22 White, 4.25 Mixed, 4.41 Asian or Asian British, 0.82 Black or Black British, 0.49 Chinese, 0.82 Other ethnic group} \\
    BMI: Mean, kg/m\(^2\)  & 28.31 (4.45) &  27.43 (4.33)  \\
    SBP: Mean, mm Hg & 147.26 (19.57) & 145.1 (18.75)\\
    DBP: Mean, mm Hg & 84.75 (10.23) & 83.22 (9.73)\\
    HbA1c: Mean, mmol/mol & 36.52 (4.32) & 36.59 (6.61)\\
    HbA1c: \% & 2.43 & 2.44 \\
    Alcohol consumption: N, P, C, NA \%  &  3.59, 5.72, 90.69, 0 &  3.92, 3.92, 91.83, 0.33\\
    \hline
  \end{tabular}
  \caption{Characteristics of patients in the \texttt{CVD+} and \texttt{CVD-} sets. N, Never. P, Previous. C, Current. NA, Not answer}
  \label{t:data_caracteristics}
\end{table}

\begin{figure}[h]
    \centering
    \includegraphics[width=1.2\textwidth]{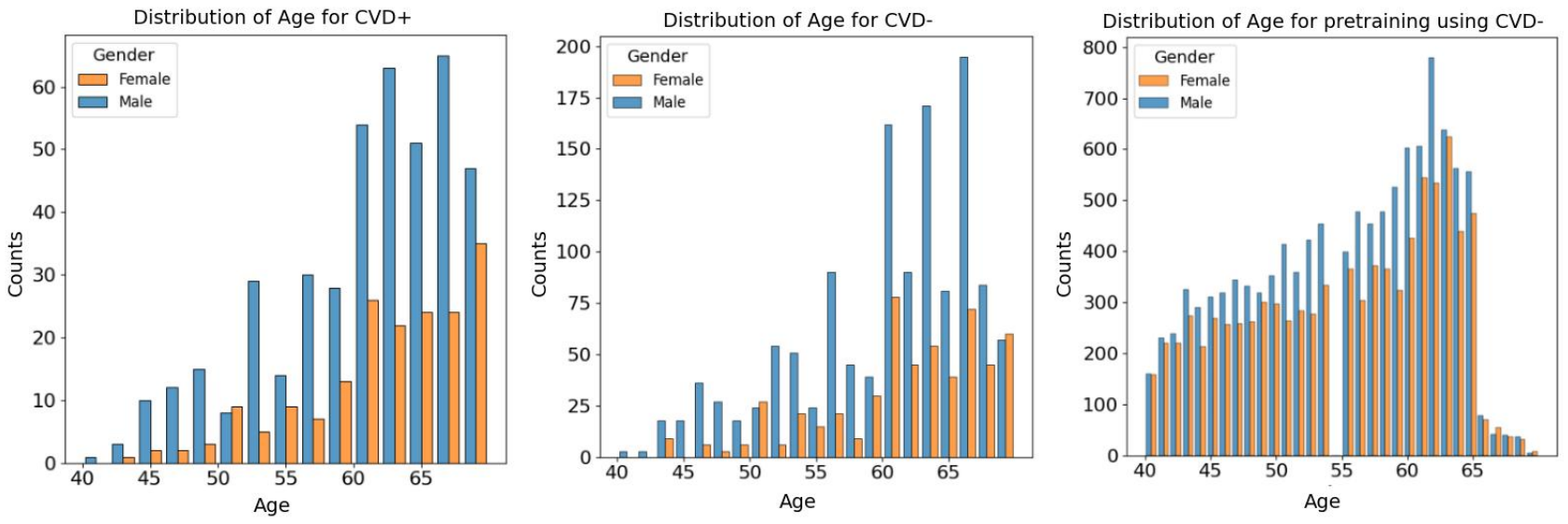}
    \caption{Distribution of age-sex cohort match. The left histogram illustrates the total of \texttt{CVD+} labeled data used solely for the classification task. The middle histogram shows the total of \texttt{CVD-} for the classification task, while the right histogram illustrates \texttt{CVD-} used for the pretraining task.}
    \label{fig:distribution_data}
\end{figure}

We used an age-sex-matched cohort  for the two groups. The relative proportion of patients included in the study in the two groups were 1:3 in the CVD and non-CVD group respectively. Utilizing an age-sex-matched study cohort offers a significant advantage by helping to mitigate the influence of confounding variables that could bias the predictive model \citep{Zhou2023AFM}. For instance, since age and sex are known to impact the risk of developing CVDs, matching groups based on these characteristics ensures that any observed differences in CVDs incidence are more directly associated with the retinal features, which are the variables of interest in this study. Previous research has indicated that matching on age and sex in the UK Biobank reduces variability within groups and enhances homogeneity among participants, thereby potentially improving the statistical power of the study \citep{Batty2019GeneralisabilityOR, Batty2020ComparisonOR}. Furthermore, machine learning models are prone to capture spurious correlations (i.e. to learn shortcuts) between predictors and targets, such as, for example, linking age or sex to the presence of pathology, unless care is taken when defining the predictors and study cohort \citep{Brown2022DetectingSL}. Figure \ref{fig:distribution_data} illustrates the age and sex in the \texttt{CVD+} and \texttt{CVD-} patient groups, which were used to train and evaluate the predictive model proposed in this study. The construction of the metadata incorporated eight clinical variables, namely sex, age, HbA1c, systolic and diastolic blood pressure, alcohol consumption, and body mass index. The decision was made to omit the smoking variable from the study analysis because a significant number of participants did not provide responses to the relevant questionnaire item. From now on, the term "metadata" will be used to describe patient details, including demographic and clinical history information.

\begin{figure}[H]
    \centering
    \includegraphics[width=\textwidth]{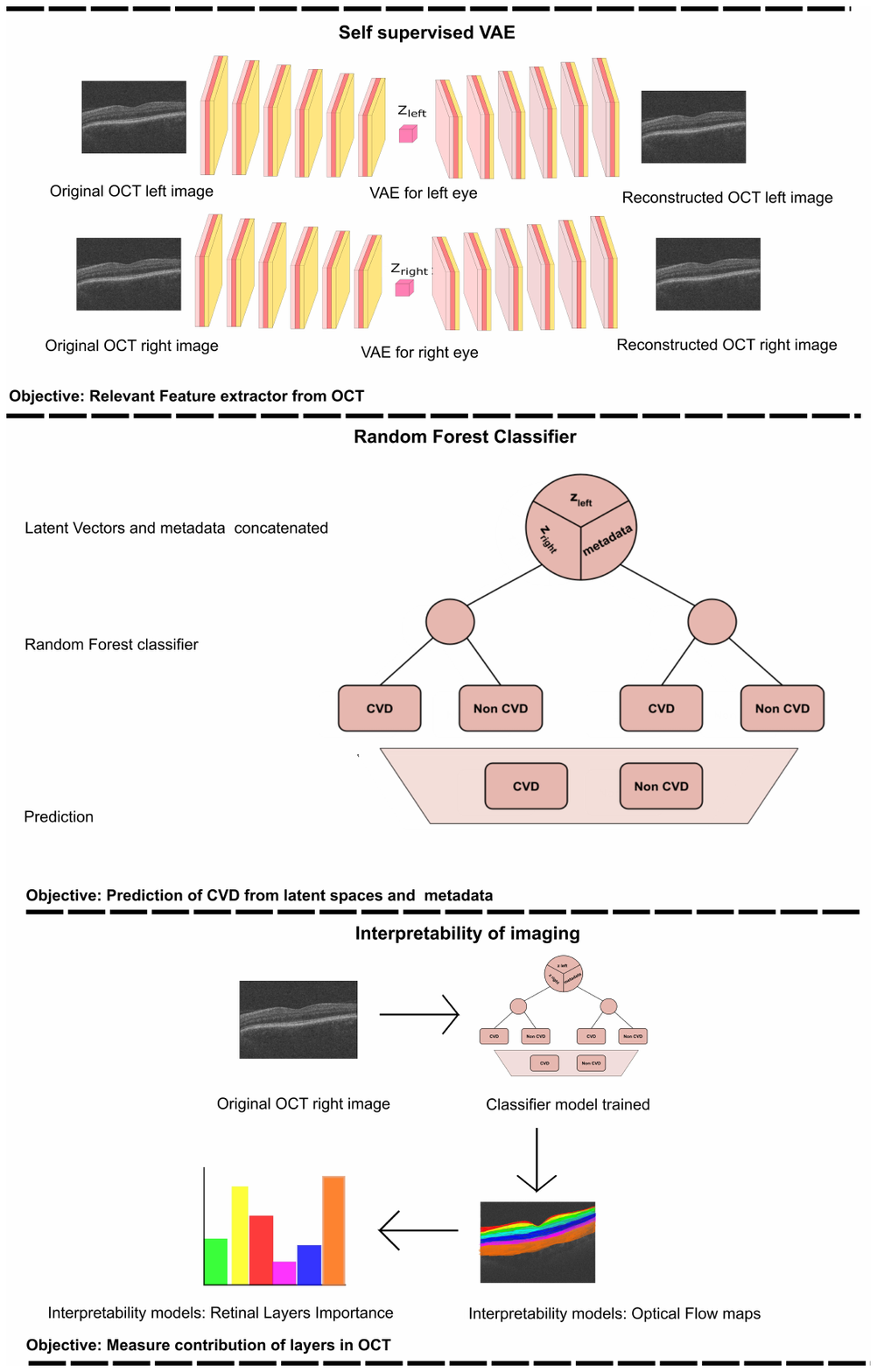}
    \caption{The provided workflow diagram illustrates the comprehensive process of training the Variational Autoencoder (VAE) and subsequently using it to acquire the latent vectors (upper section). These latent vectors are then combined with metadata and serve as inputs to the Random Forest (RF) classifier (middle section). Finally, we perform an interpretability analysis by perturbing the most relevant features, reconstructing the corresponding image and computing the optical flow between the perturbed reconstructions (lower section). $z_{left}$ represents the latent vector obtained from the training of the VAE for the left eye. $Z_{right}$ corresponds to the latent vector acquired from training the VAE for the right eye. }
    \label{fig:method}
\end{figure}

\subsection{Framework of Self-Supervised Feature Selection Variational Autoencoder and Multimodal Random Forest Classification}
This study proposes a predictive model for classifying patients into \texttt{CVD+} and \texttt{CVD-} categories, comprising a Variational Autoencoder (VAE) \citepalias{Kingma2013AutoEncodingVB} to extract features in a self-supervised manner from retinal OCT images, and a Random Forest (RF) classifier which combines the former with patient metadata and uses the resulting set of multimodal features as predictors. The proposed model consists of two stages, a self-supervised feature extraction stage, and a subsequent classification stage. A schematic diagram of the overall predictive framework is shown in Figure \ref{fig:method}.

\subsubsection{Self-supervised VAE}
In the first stage of the proposed model, a VAE is used to learn latent representations for B-scan OCT images. VAE are a widely used type of generative model in neural networks, comprising a pair of networks or network branches that are trained together, called the encoder and decoder networks/branches. Given some input data $(\mathbf{x})$, the encoder is designed to approximate the posterior distribution of the latent variables (${q}_{\phi}\left(\mathbf{z} \mid \mathbf{x}\right)$), under some assumed prior distribution $(p(\mathbf{z}))$ over the latent variables (typically, a multivariate Gaussian prior is used, that is, $p(\mathbf{z}) \sim \mathcal{N}(\cdot))$, while the decoder is trained to reconstruct the input data by sampling from the approximated posterior distribution $(p_{\theta}(\hat{\mathbf{x}} \mid \mathbf{z})$. In other words, the encoder network maps inputs to low-dimensional latent representations, and the decoder network acts as the generative model. The approximation of the true but intractable posterior distribution is obtained by maximising the lower bound of the evidence (ELBO), which can be expressed as follows:

\begin{equation}\label{eq:elbo}
\mathbf{ELBO}=\mathbb{E}_{q_{\phi}\left(\mathbf{z} \mid \mathbf{x}\right)}\left[\log p_{\theta}\left(\mathbf{x} \mid \mathbf{z}\right)\right]-D_{KL}\left(q_{\phi}\left(\mathbf{z} \mid \mathbf{x}\right) || p\left(\mathbf{z}\right)\right)
\end{equation}

The loss function utilised for training the proposed self-supervised VAE comprises two key elements: (1) the loss of mean square error (MSE) $\mathcal{L}_{MSE}$, detailed in Equation~\ref{mse_loss}, which evaluates the discrepancy in reconstruction between the original data ($\mathbf{x}_i$) and the reconstructed data ($\hat{\mathbf{x}}_i$), and (2) the loss of Kullback-Leibler divergence $\mathcal{L}_{KL}$, illustrated in Equation~\ref{kl_loss}. KL divergence quantifies the dissimilarity between the learnt latent distribution $q(\mathbf{z})$ and a previously specified distribution $p(\mathbf{z})$, which, in this scenario, the prior $p(\mathbf{z})$ is a multivariate Gaussian distribution. The parameters of the learned distribution $q(\mathbf{z})$ are its mean ($\mu_{i}$) and variance ($\sigma_{i}^{2}$). By minimising the KL divergence, the model is incentivised to shape a latent space that adheres to the target Gaussian distribution. The integration of these two loss components steers the VAE towards the dual objective of reducing reconstruction errors and aligning the learnt latent distribution with the intended prior distribution.

\begin{equation}\label{loss_function}
    \mathcal{L}_{VAE} = \mathcal{L}_{MSE} + \beta \mathcal{L}_{KL}
\end{equation}

\begin{equation}\label{mse_loss}
    \mathcal{L}_{MSE} = \frac{1}{N} \sum_{i=1}^{N} \left( \mathbf{x}_{i} - \hat{\mathbf{x}_{i}} \right) ^{2}
\end{equation}

\begin{equation}\label{kl_loss}
    \mathcal{L}_{KL} = \frac{1}{2} \left[ 1 + \log \left( \sigma_{i}^{2}\right) - \sigma_{i}^{2} - \mu_{i}^{2} \right]
\end{equation} 

We trained independent VAEs for each eye, to learn unique latent features from the OCT images. Subsequently, a classifier used these learnt features to predict the probability of an individual's prospective CVD incidence. 

\subsubsection{Classification} 
Using the features acquired from the VAE in the previous stage, we trained a Random Forest (RF) classifier to distinguish between individuals in the \texttt{CVD+} and \texttt{CVD-} categories, as illustrated in Figure \ref{fig:method}. The input for this process consists of the latent vector representation of each OCT image generated by the VAE for each eye, which is merged with a vector containing the relevant patient information. RF are a type of ensemble machine learning method that involves multiple decision trees, each of which is trained on a randomly selected subset of training data \citep{Breiman2001RandomF}. Using the power of numerous decision trees and incorporating random feature selection, this ensembling technique enhances the generalisability of the predictive model to new data by reducing model variance by averaging predictions from the trees in the ensemble. RF have been widely applied in medical settings for both classification and regression tasks, including in previous studies related to CVD diagnosis \citep{khozeimeh2022rf,yang2020study}. One notable advantage of RF compared to other classification algorithms is their ability to easily handle multimodal data that include various data types (such as categorical, ordinal, and continuous). Decision trees within the ensemble operate independently, and their combined predictions are aggregated to produce the final RF prediction for a given input using majority voting for classification tasks. This structure also provides feature importance, which enhances the explainability of the model's decisions. Additionally, RF is computationally efficient when compared to more complex models, such as neural networks, as it does not require GPU resources for training.


\subsection{Experiment Details}
\label{exp}

All experiments were carried out with an NVIDIA Tesla M60 GPU. The model was trained using PyTorch (v1.10.2) and a grid search strategy was used with five-fold cross-validation to determine the best hyperparameters. The data set was divided into training, validation, and test sets in a ratio of 6: 2: 2. The encoder and decoder networks were constructed with six 2D convolution layers each; the encoder used Rectified Linear Unit (ReLU) activations, while Leaky Rectified Linear Unit (LeakyReLU) activations were used in the decoder (see the upper section of Figure \ref{fig:method}). The training of our model occurred in two phases, initially engaging in self-supervised learning for each eye independently to acquire latent representations, which were then utilised to initialise the subsequent fine-tuning phase.

During the classification phase, we conducted a thorough investigation into the impact of latent representations derived from the OCT images of the left and right eyes, along with patient metadata, on the predictive task. This was accomplished by generating seven datasets from the same group of 2846 patients, each comprising different combinations of data sources: (i) latent representations from the left eye only (LE); (ii) latent representations from the right eye only (RE); (iii) latent representations from both eyes (BE); (iv) metadata only (MTDT); (v) left eye with metadata (LE-MTDT); (vi) right eye with metadata (RE-MTDT); and (vii) both eyes with metadata (BE-MTDT). Random Forest classifiers were trained separately on each of these seven datasets, as depicted in Figure \ref{fig:method} for the BE-MTDT dataset.
Finally, the optimal hyperparameter values of RF classifiers were determined through a combination of grid search and empirical experimentation, to identify the best performing RF model for each specific dataset. We divided the dataset into training, validation, and test sets, following a distribution ratio of $\sim 5:2:3$, respectively (resulting in, 1423 patients in the training set, 459 patients in the validation set and 964 patients in the held-out unseen test set). Grid search was performed using five-fold cross-validation, while an independent, unseen test set remained fixed throughout all experiments to evaluate all trained classifiers fairly. During the optimal hyperparameter search, we employed Recursive Feature Elimination (RFE) as a feature selection technique to address overfitting and train the model with the most pertinent variables for classification. In the majority of cases, the model was trained with the top 10 most significant features, with the exception of the RE case, where the optimal conditions involved using only 5 variables. 

To evaluate the effectiveness of our model, we compared predictive performance against the QRISK3 algorithm, the current gold standard used by healthcare professionals / cardiologists to assess the patient's risk of stroke or heart attack, in a 10-year period. The QRISK3 score was calculated within the specified test dataset, following the methodological guidelines outlined in \citep{Li2019RP}. The evaluation of the QRISK3 score involved entering essential variables which are described in Table \ref{t:qrisk_variables}. In cases where certain information was missing, we consistently represented these gaps as '0' when evaluating the QRISK3 scores on the test dataset. For our classification task, we evaluated the model performance using a range of metrics. Accuracy, sensitivity, and specificity were determined by calculating true positives, true negatives, false positives, and false negatives (using a classification probability threshold of $t=0.5$, i.e. if the predicted probability is $\geq 0.5$, the patient is classified as \texttt{CVD+}, else as \texttt{CVD-}). The receiver operating characteristic (ROC) curve was constructed by computing the true positive rate and false positive rate. The area under the ROC curve (AUROC) was then employed as a performance measure to assess both our model and the QRISK3 algorithm.

\subsection{Model Explainability}
\label{mexp}
Predictive models based on machine/deep learning algorithms, proposed for identifying risk of disease from medical imaging often fail to report both 'local' and 'global' explanations for the model's predictions. This is especially prevalent in the case of deep learning-based approaches that are often treated as black boxes, with little information provided on the mechanism by which models arrive at specific predictions. Local explanations provide insights to individual decisions/predictions of the model. For example, this may involve identifying specific input variables/regions of an image that had the most influence on the model's prediction for that instance. On the other hand, global explanations describe the model's behaviour across predictions for all instances in all classes of interest. Specifically, global explanations provide information on the most common discriminative features identified by the model for all instances in each class of interest. Providing both local and global explanations of model behaviour is essential for developing responsible AI in healthcare applications, as it can help identify systematic biases in data and mitigate for the same (e.g., learning of 'short-cuts' is a common issue encountered in the application of deep learning-based methods for predictive tasks using medical images), build trust in AI systems by improving transparency in model decision making, and may even provide new insights to previously known associations between image-derived phenotypes and the presence or progression of diseases. Therefore, in this study, we employ distinct techniques to provide both local and global explanations for the proposed predictive model.

To provide local explanations of the behaviour of the model and elucidate how the model uses OCT imaging-derived features to classify instances in the \texttt{CVD+} group, we first selected the best performing RF classifier according to all classification metrics used to evaluate and compare all investigated models. Subsequently, based on the selected RF classifier, we identified the latent variable derived from the OCT image with the highest importance assigned by the RF, which we denoted $\mathbf{z}_{max}$. To visually assess the regions of the retina in OCT images that contribute significantly to the prediction of CVD, we propose a novel optical flow-based latent traversal approach that evaluates the impact of perturbing the most important latent feature $\mathbf{z}_{max}$ on subsequent reconstructions of OCT images. 
Specifically, given an image \textbf{x}, we compute the corresponding latent vector $\mathbf{z}$ using the trained encoder network. Next, we perturb only the dimension $\mathbf{z}_{max}$ of the calculated latent vector and reconstruct the corresponding image. This perturbation is performed by multiplying the latent dimension $\mathbf{z}_{max}$ by the standard deviation of this latent component calculated throughout the training population, defined as $\sigma_{max}$. The remaining latent variables in the computed latent vector are left unchanged, resulting in a perturbed latent vector ($\mathbf{\hat{z}}$). Finally, we reconstruct the input OCT image ($\mathbf{\hat{x}}$) using the perturbed latent vector $\mathbf{\hat{z}}$.

To visualise the regions in the reconstructed OCT images affected by the altered latent vector $\mathbf{\hat{z}}$, we examine the variances between the initial image $\mathbf{x}$ and the reconstructed image $\mathbf{\hat{x}}$ derived from $\mathbf{\hat{z}}$. In this context, we calculate the optical flow between these images using the Lucas-Kanade algorithm \citepalias{Lucas1981AnII}. The resulting optical flow, showing the magnitude of the displacement vector for the moving pixels, was then superimposed on the original image to create a visual representation. The optical flow forms a vector field that indicates how the pixels between the images (that is, $\mathbf{x}$ and $\mathbf{\hat{x}}$) change due to the latent traversal from $\mathbf{z}$ to $\mathbf{\hat{z}}$. The estimated vector field between $\mathbf{x}$ and $\mathbf{\hat{x}}$ helps to visually illustrate the regions in the OCT image that were altered by modifying the latent component $\mathbf{z}_{max}$. This aids in pinpointing the areas of the image influenced by alterations to the crucial latent variable for accurately classifying a patient's CVDs risk based on their OCT image(s), and consequently, helps to understand which retinal areas are informative for distinguishing between the \texttt{CVD+} and \texttt{CVD-} patient groups.

To provide global explanations of the behaviour of the model, we calculate the importance of the characteristics assigned to each characteristic by the RF in each predictive model investigated. As mentioned previously, the RF in each predictive model were trained using a reduced set of characteristics identified by RFE. Feature importance is calculated in RF as the average Gini information gain for any given feature, calculated across all decision trees in the forest. The feature importance values for all classifiers studied in this work, across all test set instances, are summarized as bar plots later. Additionally, we calculate the relative importance of the type / channel of data used as inputs/predictors in this study, namely, OCT images of the left and right eye and patient metadata, for the task of distinguishing between the \texttt{CVD+} and \texttt{CVD-} groups.


\section{Results}

\subsection{Classification Performance}
As discussed previously, we trained and evaluated the performance of several RF classifiers, where each classifier was trained and evaluated independently using seven different combinations of data types obtained from the same set of patients (comprising \texttt{CVD+} and \texttt{CVD-} groups). Specifically, the datasets used were LE, RE, BE, LE-MTDT, RE-MTDT, BE-MTDT and MTDT. Henceforth, for brevity, we refer to classifiers trained and evaluated on these datasets as LE-RF, RE-RF, BE-RF, LE-MTDT-RF, RE-MTDT-RF, BE-MTDT-RF, and MTDT-RF. The performance of all seven classifiers was evaluated and compared using the same unseen test set (which contains 964 patient data, 834 \texttt{ CVD-} and 130 \texttt{CVD+}), and using the same set of evaluation metrics outlined in Section~\ref{exp}. The rationale for comparing all seven classifiers against each other was to - (i) assess whether combining latent features learnt from OCT images of both eyes (BE) provided greater discriminative power than using those from either left (LE) or right eye (RE) alone; (ii) compare the discriminative power of OCT image-derived latent features against patient metadata; and (iii) evaluate the discriminative power gained by enriching OCT image-derived latent features with patient metadata. 

The performance of all seven classifiers on the unseen test set is summarised in Table \ref{t:metrics}. These results show that the BE-MTDT-RF classifier consistently outperformed all six other classifiers, with statistically significant differences in p-values (refer to Table \ref{t:chisquared_all}). This suggests that combining information from OCT images of both eyes (ie, learnt latent representations) with patient metadata was more informative in distinguishing between the \texttt{CVD+} and \texttt{CVD-} groups. In terms of evaluating the effectiveness of OCT image-derived characteristics and metadata information for classifying \texttt{CVD+} and \texttt{CVD-} patients, results for the BE-MTDT-RF, LE-MTDT-RF, RE-MTDT-RF and MTDT-RF classifiers indicate that combining latent features learnt from OCT images with patient metadata (i.e. BE-MTDT-RF, LE-MTDT-RF, RE-MTDT-RF), consistently improves classification performance relative to using patient metadata alone (MTDT-RF). Specifically, the BE-MTDT-RF classifier demonstrated the highest performance, achieving an accuracy of 0.70, sensitivity of 0.70, specificity of 0.70, and an AUC score of 0.75. Furthermore, combining latent features from RE or LE OCT images with metadata (i.e., LE-MTDT-RF and RE-MTDT-RF) resulted in improvements of $13-17\%$ across all classification metrics relative to the MTDT-RF classifier. Notably, the MTDT-RF classifier exhibited the lowest values across all classification metrics, with an accuracy, sensitivity, and specificity of 0.52, and an AUC score of 0.57. The BE-RF, RE-RF, and LE-RF classifiers also consistently outperformed the MTDT-RF classifier in all classification metrics. However, they were outperformed by classifiers that incorporated OCT image-derived features with patient metadata (i.e., BE-MTDT-RF, LE-MTDT-RF, RE-MTDT-RF).

\begin{table}[H]
  \centering
  \renewcommand{\arraystretch}{1.4}
  \begin{tabular}{ |p{3cm}|p{2cm}|p{2cm}|p{1.5cm}|p{1.5cm}|}
    \hline
    \multicolumn{5}{|c|}{Classification metrics} \\
    \hline
    Classifiers & Accuracy & Sensitivity & Specificity & AUC \\
    \hline
    LE-MTDT-RF & 0.68 & 0.69 & 0.68 & 0.72 \\
    RE-MTDT-RF & 0.67 & 0.69 & 0.67 & 0.70 \\
    \textbf{BE-MTDT-RF} & \textbf{0.70} & \textbf{0.70} & \textbf{0.70} & \textbf{0.75} \\
    MTDT-RF & 0.52 & 0.52 & 0.52 & 0.57 \\
    LE-RF & 0.61 & 0.56 & 0.62 & 0.64 \\
    RE-RF & 0.58 & 0.53 & 0.59 & 0.62\\
    BE-RF & 0.64 & 0.57 & 0.65 & 0.67\\
    \hline
  \end{tabular}
  \caption{Predictive analysis of cardiovascular disease (CVD) metrics utilizing UK Biobank data across seven distinct cases.}
  \label{t:metrics}
\end{table}

\begin{figure}[H]
    \centering
    \includegraphics[width=\linewidth]{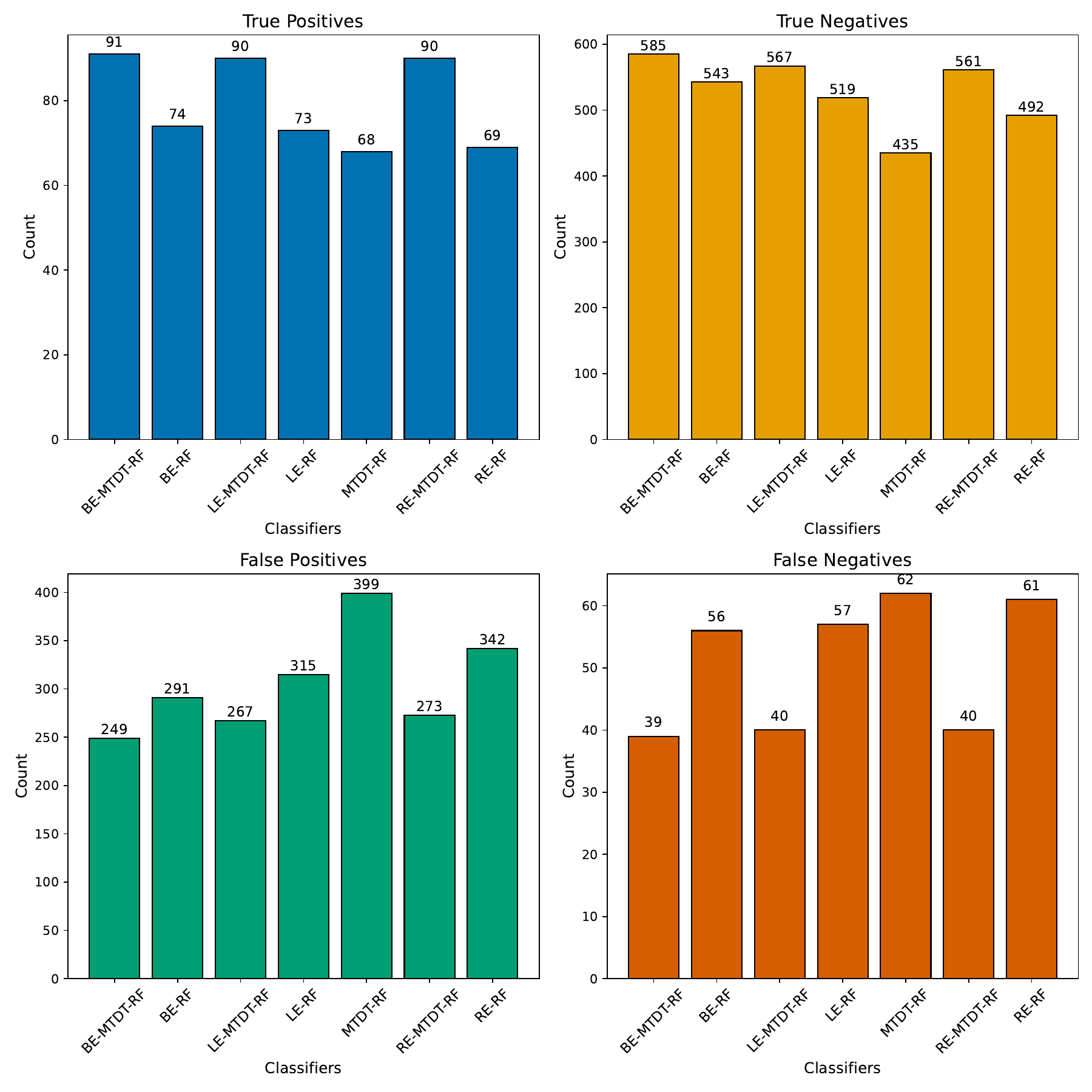}
    \caption{Comparison of classifier performance in terms of True Positives (TP), True Negatives (TN), False Positives (FP), and False Negatives (FN). Values above the bars represent the corresponding counts for each classifier.}
    \label{fig:cm_all}
\end{figure}

Figure \ref{fig:cm_all} presents four histograms depicting true positives, true negatives, false positives, and false negatives for the seven classifiers investigated. A consistent observation across all our results is that the BE-MTDT-RF classifier misclassified fewer instances in the \texttt{CVD+} (39 out of 130) group, than all other classifiers, which is consistent with the classification metrics summarised in Table~\ref{t:metrics}. Similarly, the LE-MTDT-RF and RE-MTDT-RF classifiers exhibited good sensitivity by incurring few false negative errors, i.e. 40 cases out of 130 instances in the \texttt{CVD+} group were incorrectly classified. The MTDT-RF classifier yielded fewer true positives and true negatives (68 and 435, respectively) compared to the other classifiers that utilized only OCT features, which aligns with the results presented in Table~\ref{t:metrics}.

\begin{figure}[H]
    \centering
    \begin{subfigure}[b]{0.45\textwidth}
        \includegraphics[width=\linewidth]{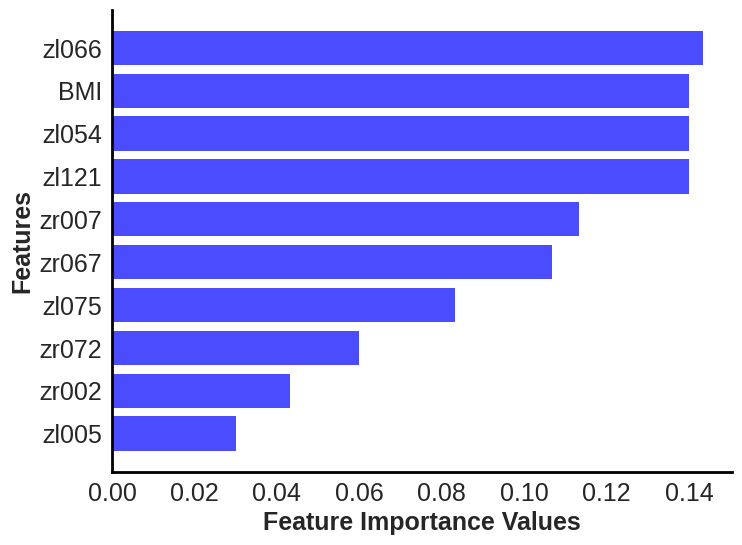}
        \caption{BE-MTDT-RF}
        \label{fig:features_be_mtdt}
    \end{subfigure}
    \hfill
    \begin{subfigure}[b]{0.45\textwidth}
        \includegraphics[width=\linewidth]{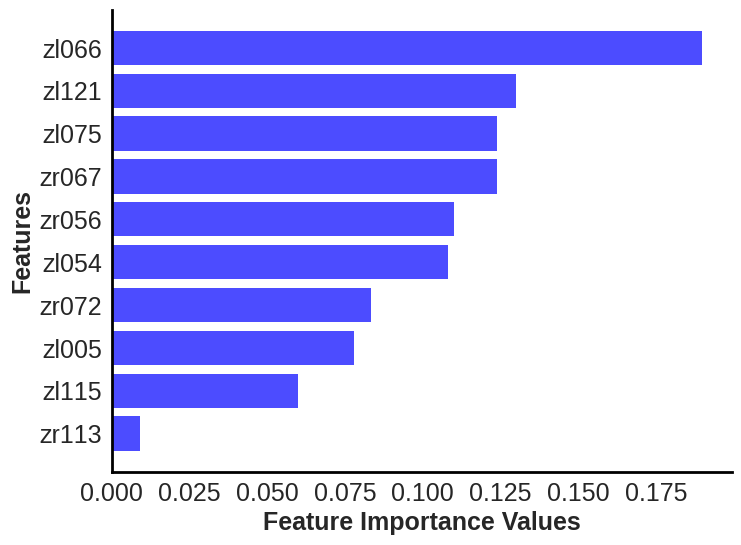}
        \caption{BE-RF}
        \label{fig:features_be}
    \end{subfigure}
    
    \begin{subfigure}[b]{0.45\textwidth}
        \includegraphics[width=\linewidth]{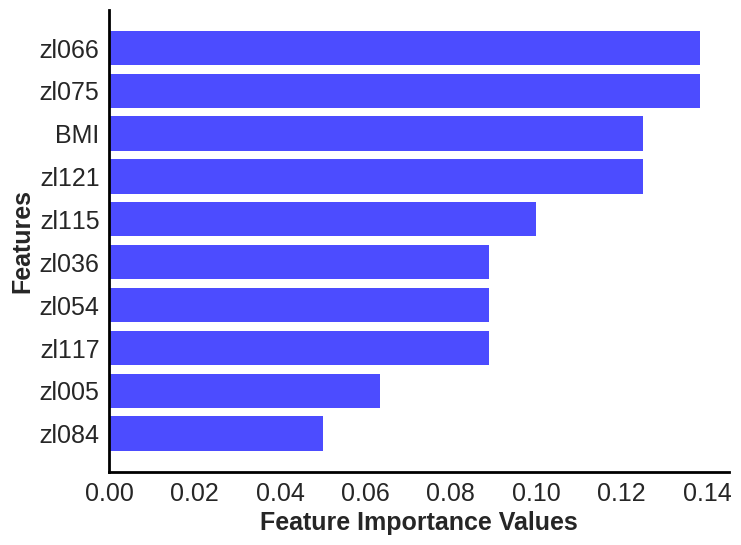}
        \caption{LE-MTDT-RF}
        \label{fig:features_le_mtdt}
    \end{subfigure}  
    \hfill
    \begin{subfigure}[b]{0.45\textwidth}
        \includegraphics[width=\linewidth]{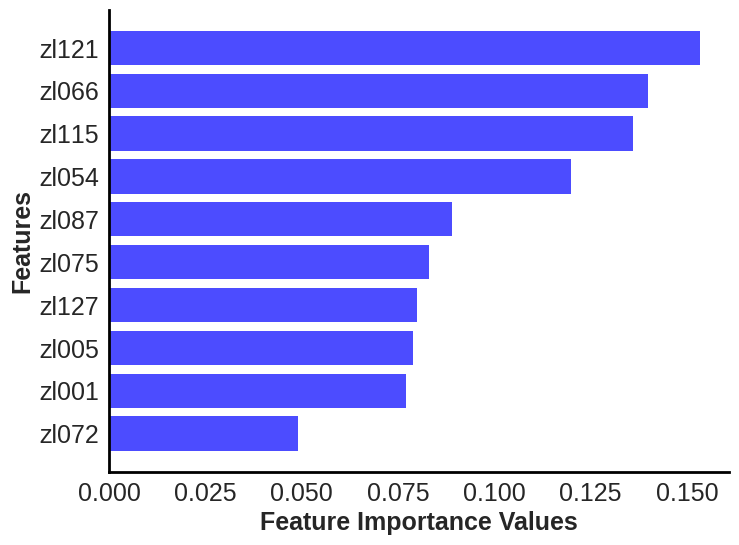}
        \caption{LE-RF}
        \label{fig:features_le}
    \end{subfigure}   
    
    \begin{subfigure}[b]{0.45\textwidth}
        \includegraphics[width=\linewidth]{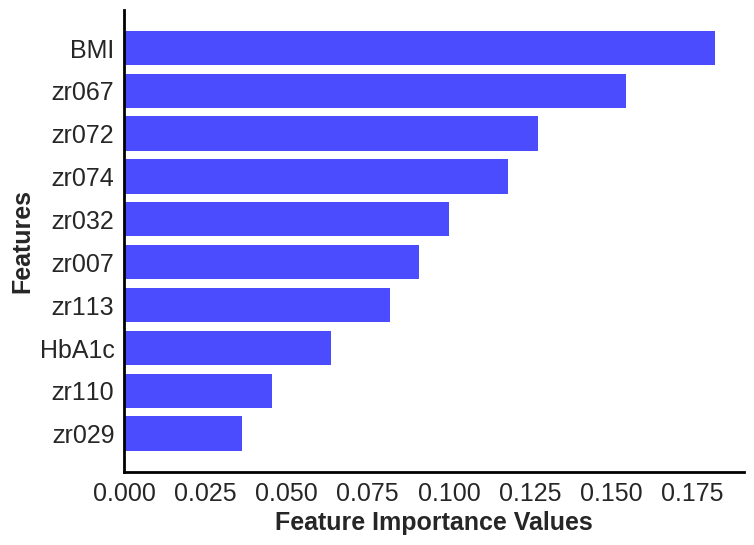}
        \caption{RE-MTDT-RF}
        \label{fig:features_re_mtdt}
    \end{subfigure}  
    \hfill
    \begin{subfigure}[b]{0.45\textwidth}
        \includegraphics[width=\linewidth]{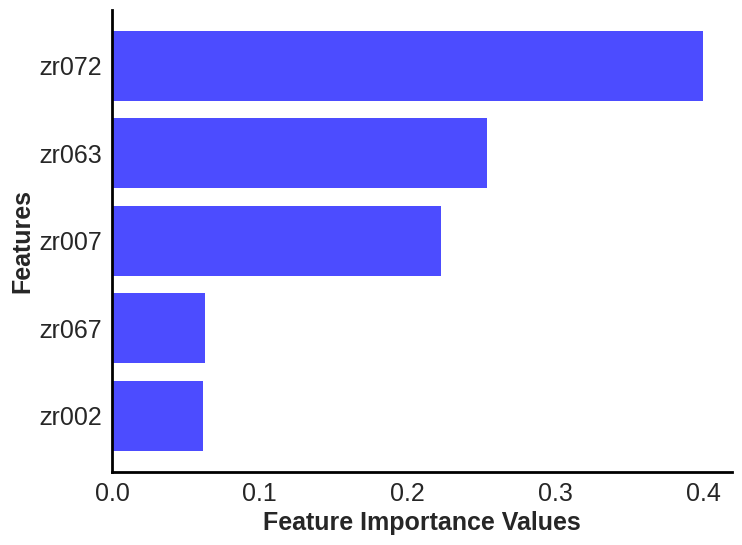}
        \caption{RE-RF}
        \label{fig:features_re}
    \end{subfigure}
    
    \begin{subfigure}[b]{0.45\textwidth}
        \includegraphics[width=\linewidth]{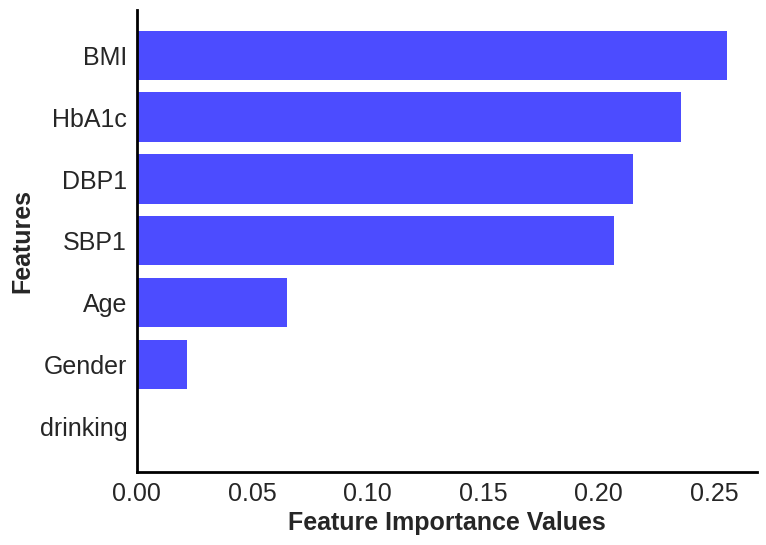}
        \caption{MTDT-RF}
        \label{fig:features_mtdt}
    \end{subfigure}  
    
    \caption{Calculation of feature importance magnitudes for the seven different classifiers investigated, where each classifier uses different combinations of data channels/modalities.}
    \label{fig:features_importances}
\end{figure}

A significant observation in the results is that including both eyes was advantageous for both cases, BE-MTDT-RF and BE-RF, compared to their counterparts, LE-MTDT-RF and RE-MTDT-RF, RE-RF, and LE-RF, respectively. Furthermore, in both scenarios, with/without metadata, the left eye consistently provided improved classification performance compared to the right eye. This finding was statistically significant, as indicated by the p-values reported in Table \ref{t:chisquared_left_right}. This observation is in concordance with the global explanations of models' predictions summarised in Figure \ref{fig:features_importances}, wherein features attributed to the left eye were found to be more discriminative (i.e., had higher feature importance) than those of the right eye. We posit that the impact on the latent vectors associated with the left eye is interconnected with the superior image quality of the images of the left eye within our cohort (see Figure \ref{fig:qis}). The UKBB standard operating procedure stipulated that the second eye imaged was consistent with the left eye. This protocol was not randomised. As a result, there may be potential systematic disparities between left and right eye OCTs (for example, left eye scans might consistently exhibit better quality because they are the second scan performed and patients are potentially more adept at following instructions). These collective findings underscore the improved performance achieved by integrating retina OCT imaging and metadata in the classification task.

Subsequently, we conducted a rigorous comparative analysis between the best classifier identified from the previous experiments, namely BE-MTDT-RF, and the QRISK3 algorithm, the current clinical standard for assessing patients at risk of stroke or MI. As elucidated in Table \ref{t:metrics_qrisk}, our model showed superior classification performance in terms of accuracy, sensitivity, apecificity and AUC. Specifically, for the QRISK3 model, performance metrics were as follows: accuracy (0.55), sensitivity (0.6), specificity (0.545), and AUC (0.60). In contrast, BE-MTDT-RF achieved the following metrics: accuracy (0.70), sensitivity (0.70), specificity (0.70), and AUC (0.75). This discrepancy in performance at separating the \texttt{CVD+} and \texttt{CVD-} patient groups underscores the improved discriminative capacity provided by our model which combines retinal OCT imaging data with basic patient information (see Table~\ref{t:data_caracteristics}), relative to using an extensive set of demographic and clinical variables (refer to Table \ref{t:qrisk_variables} for more information on the variables of the patients required to calculate the QRISK3 scores), as in the case of the widely used QRISK3 algorithm.

\begin{table}[H]
  \centering
  \renewcommand{\arraystretch}{1.4}
  \begin{tabular}{ |p{2.5cm}|p{2cm}|p{2cm}|p{2cm}|p{2cm}|}
    \hline
    \multicolumn{5}{|c|}{Classification metrics} \\
    \hline
    Modality & Accuracy & Sensitivity & Specificity & AUC-Value \\
    \hline
    \textbf{BE-MTDT-RF} & \textbf{0.70} & \textbf{0.70} & \textbf{0.70} & \textbf{0.75} \\
    QRISK & 0.55 & 0.60 & 0.545 & 0.60\\
    \hline
    \multicolumn{5}{l}{\small{$\chi^2 = 95.72$, $df=1$, $p=1.31 \times 10^{-22}$}} \\
  \end{tabular}
  \caption{Comparison of classification metric results between our model employing both ocular data and metadata (BE-MTDT) and the QRISK algorithm. McNemar’s Test (*$p < 0.005$)}
  \label{t:metrics_qrisk}
\end{table}

\subsection{Model Explainability}

In order to provide local explanations for the behavior of all classifiers investigated in this study, we analysed the most important features identified by each model (refer to Figure~\ref{fig:features_importances}) for distinguishing between the \texttt{CVD+} and \texttt{CVD-} groups. Important features identified for the best performing classifier, namely, BE-MTDT-RF in particular, provided some noteworthy insights. As highlighted in Figure~\ref{fig:features_be_mtdt}, we found that a latent variable learned from the left-eye OCT image, denoted zl066, had the most influence on the classifier's ability to separate \texttt{CVD+} and \texttt{CVD-} patient groups. Additionally, among the top 10 most important features identified for the BE-MTDT-RF classifier, 9 of the features pertained to latent variables learned from the left-eye OCT image. BMI was the only feature from the basic set of patient metadata used to train the classifier, that was identified to have a significant influence on the classifier's predictions. Furthermore, looking at the local explanations summarised in Figure~\ref{fig:features_be}, \ref{fig:features_le_mtdt}, \ref{fig:features_le}, we observe that latent variable zl066 consistently ranks among the top two most important features for the BE-RF, LE-MTDT-RF, and LE-RF classifiers, respectively. This indicates that the retinal features encoded by zl066 are consistently considered to be relevant across all four classifiers presented in Figure~\ref{fig:features_be_mtdt},\ref{fig:features_be}, \ref{fig:features_le_mtdt}, \ref{fig:features_le}. Among the classifiers which combined retinal OCT-image derived features with patient metadata, namely, BE-MTDT-RF, LE-MTDT-RF and RE-MTDT-RF, we observed that only two features, namely, BMI and HbA1c, ranked among the top 10 most important features for the classification task. Both features are known and established cardiovascular risk factors, and importantly, we infer from these results that the latent variables learned from the retinal OCT images, had a greater influence on the classifiers' predictions than the patient metadata variables. As previously highlighted, we hypothesize that the significant importance of the latent variables corresponding to the left eye can be attributed to the superior image quality of left-eye OCT images within our cohort (as illustrated in Figure \ref{fig:qis} in the Appendix). As a result, corresponding latent vectors $\mathbf{z}$ effectively capturing image features that potentially enhance predictive capabilities. 

\begin{figure}[H]
    \centering
    \includegraphics[scale=0.33]{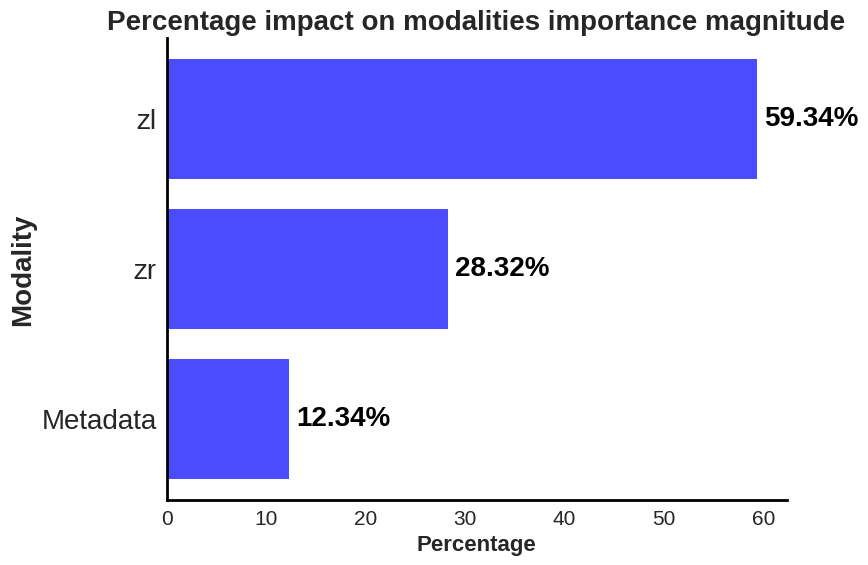}
    \caption{Global explanations of features from different data modalities/channels which were considered important by the predictive model for separating the \texttt{CVD+} and \texttt{CVD-} groups. Bar plot summarises the relative importance of latent variables from left $(zl)$ and right $(zr)$ eye OCT images and patient metadata, as percentages.}
    \label{fig:total_importances}
\end{figure}

\begin{figure}[H]
    \centering
    \includegraphics[width=0.95\textwidth]{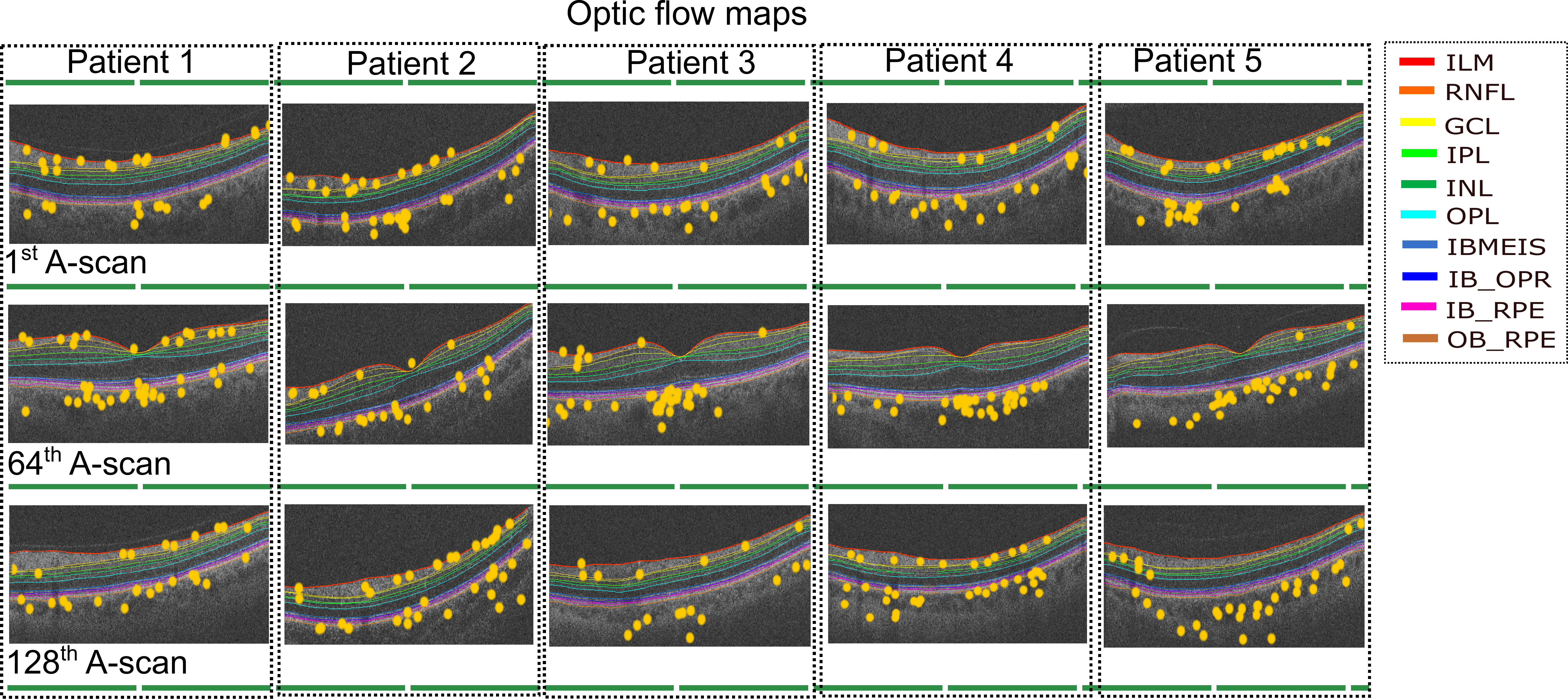}
    \caption{Optical flow maps presented for three differents B-scans for the left eye. The top row corresponds to the 1st B-scan, the middle image for the 64th B-scan while the bottom row depicts the final B-scan. The yellow circles represent the regions that the optical flow maps highlight when modifying the latent variable zl066. The legend includes the names of layer boundaries in the optical coherence tomography images, as follows:  Internal Limiting Membrane; RNFL: Retinal Nerve Fiber Layer; GCL: Ganglion Cell Layer; IPL: Inner Plexiform Layer; INL: Inner Nuclear Layer; OPL: Outer Plexiform Layer;  BMEIS:	Boundary of Myoid and Ellipsoid of Inner Segments; IB\textunderscore OPR:	Inner Boundary of Outer Segment Retinal Pigment Epithelium Complex; IB\textunderscore RPE: Inner Boundary of Retinal Pigment Epithelium; OB\textunderscore RPE: Outer Boundary of Retinal Pigment Epithelium}
    \label{fig:optical_flow}
\end{figure}

Using the insights gained from analysing the global explanations of classifier behavior summarised in Figure~\ref{fig:total_importances}, we propose a novel approach based on latent space traversals to translate the former into local explanations that provide insights to regions of the OCT image that contain relevant information for correctly identifying patients at risk of cardiovascular disease. Specifically, having identified latent variable zl066, derived from left-eye OCT images, as being the most important feature for classification, our local explainability approach (refer to section \ref{mexp}) begins by perturbing the values of the latent variable for any image in the \texttt{CVD+} group, reconstructs the OCT image using the perturbed latent representation (using the pretrained VAE) and then estimates optical flow maps between the original and perturbed OCT image reconstructions, seeking to pinpoint the specific image regions that change as a result of the perturbation. 
We conducted a qualitative analysis involving estimation of optical flow maps between the original reconstructed B-scan OCT images and their perturbed counterparts. This analysis was specifically focused on representative B-scan examples from 5 patients correctly diagnosed with CVD, where each patient corresponds to a column in Figure \ref{fig:optical_flow}. We considered 3 B-scans per patient, including the 1st one (upper row), 64th (middle row) and 128th (bottom row). Finally, by overlaying the estimated optical flow maps onto the original OCT images, we obtained visual interpretations of the retinal image features encoded by the latent variable of interest.

The optical flow maps generated by our model predominantly accentuated the choroidal layer in the majority of B-scans, with additional identification of layers adjacent to the choroid, including the retinal pigment epithelium (RPE). Additionally, the optical flow maps highlighted the regions in the inner retinal layers, likely corresponding to the retinal nerve fiber layer (RNFL) and ganglion cell layer (GCL) \citep{Zhou2023AFM}. Although some other layers received comparatively less emphasis, the main focus remained on the choroidal layer and the innermost layers. The optical flow maps provide precise localisation of the image regions modelled by latent variable zl066 (visualised as landmarks, as shown in Figure~\ref{fig:optical_flow}), thereby providing local explanations for the most discriminative regions within the OCT, and providing insights to which retinal layers may contain relevant information for predicting risk of CVD in patients. In particular, these local explanations highlighted the relevance of information contained within the choroidal layer of the retina, for distinguishing between \texttt{CVD+} and \texttt{CVD-} patient groups. Such findings have significant implications for understanding the potential association between changes to the choroidal layer in the retina and the onset and progression of cardiovascular diseaes.


\section{Discussion}
Our findings indicate that the use of retinal OCT images in conjunction with VAE and multimodal RF classification has potential to improve the ability to identify patients at risk of CVDs (within a five-year interval), relative to the use of the current clinical standard, that is, the QRISK3 score. This approach can serve as a complementary strategy to existing clinical procedures used for the prevention of primary CVDs \citep{Badawy2022EvaluationOC}. Our investigation included the deployment of a self-supervised VAE coupled with an RF classifier framework, which incorporates B-scan OCT images and metadata as distinct modalities. This integration allowed our model to discern the specific attributes within the OCT images that contribute significantly to the prediction of CVDs. Importantly, our study distinguishes itself by interpreting the particular OCT image features (at both the global i.e. class/category, and local i.e. instance, levels), which are relevant to the classification task and thereby provide insights to the key regions of the retinal image that are most discriminative. To the best of our knowledge, some studies have ventured into the application of OCT within a primary care framework for CVDs. However, these studies were limited in their explanatory capacity regarding the effects of including images from both eyes and different types of patient data, and used a small portion of the OCT B scans, limiting the information from the entire volume. Nevertheless, the performance results show promising outcomes for OCT as a modality in the primary care of CVDs \citep{Garca2022PredictingMI, Zhou2023AFM}.

Interestingly, our results suggest that choroidal morphology is a predictor of identifying patients at risk of CVDs, which is consistent with previous studies \citep{YEUNG2020473} that have reported significant associations between choroidal characteristics and the risk of stroke and acute myocardial infarction. Given that the choroid has the highest flow per perfused volume of any human tissue and that there is growing evidence that changes in the choroid microvasculature can be indicative of systemic vascular pathology \citep{Ferrara2021BiomechanicalPO}, our findings offer clinical interpretability to the predictions of our classifier. Currently, UK Biobank images are captured using a spectral domain (SD) OCT \citep{Keane2016OpticalCT}. SD-OCT images suffer significant light scattering at the choroid, which limits the resolution of this layer. Despite this limited resolution, it is encouraging to observe that the proposed approach focused on features within the choroidal layer to identify patients at risk of stroke or myocardial infarction. 
To provide a greater context to the key findings reported in this study, it appears probable that once image modalities with deeper tissue penetration, such as swept source OCT, become available at scale in population imaging initiatives (such as UK Biobank), the predictive performance of learning-based systems such as ours will improve \citep{Copete2013DirectCO}. In addition to the choroidal characteristics, our results indicated that the inner retinal layers contributed to the classification, including the retinal nerve fibre layer (RNFL), the ganglion cell layer (GCL), and the inner plexiform layer (IPL). These aspects of the neurosensory retina consist of retinal ganglion cells, their synapses with bipolar cells, and their axons.  Regarding the choroid, thinning and defects in these layers have received extensive study in relation to established CVDs, but their role as predictors of future disease has received limited attention \citep{Chen2023RetinalNF, Matuleviit2022RetinalAC}. Mechanisms that may underpin the role of neurosensory retina morphology as a predictor of CVDs are yet to be elucidated, although it could be hypothesised that subclinical ocular circulatory pathology could explain morphological changes through local ischaemic damage \citep{Chen2023RetinalNF}, or neuronal degeneration could even occur through silent / subclinical cerebral ischaemic / vascular changes manifesting in the inner retina through transneuronal retrograde degeneration \citep{Langner2022StructuralRC, Park2013TransneuronalRD}.

\section{Conclusion}
This study presents a predictive framework comprising a self-supervised representation learning approach based on a VAE and a Random Forest classifier, which effectively integrates multi-modal imaging (OCT imaging) and non-imaging (e.g. patient demographic and clinical variables) data, to identify patients at risk of stroke or myocardial infarction. Although the focus of this study was on spectral domain OCT imaging, future improvements to the presented work could include the use of more informative retinal imaging modalities such as swept source OCT imaging or wide-field OCT angiography (OCTA) imaging. We hypothesise that learning representations from multi-modal retinal imaging (e.g. fundus photographs, OCT, OCTA) may improve the classification performance of the proposed approach. Additionally, a key benefit of the proposed approach is that it lends itself to explaining model predictions in both a global (across all instances) and local (instance-specific) sense, and thereby, provides insight into which retinal layers contain the most relevant information to identify risk of CVDs. In general, this investigation has supported the utility and prospective predictive value of OCT imaging to identify people at risk of stroke or myocardial infarction. As OCT imaging is a cost-effective and noninvasive imaging modality, the results presented in this study make a compelling case for future exploration of OCT imaging as a tool for screening individuals at risk of cardiovascular disease.

\onecolumn


\makeatletter
\renewcommand \thesection{S\@arabic\c@section}
\renewcommand\thetable{S\@arabic\c@table}

\renewcommand{\thesubfigure}{S\@arabic\c@figure\alph{subfigure}}
\renewcommand \thefigure{S\@arabic\c@figure}
\makeatother

\setcounter{section}{0}
\section{Supplementary Figures and Tables}

\begin{figure}[H]
    \centering
    \includegraphics[width=0.99\textwidth]{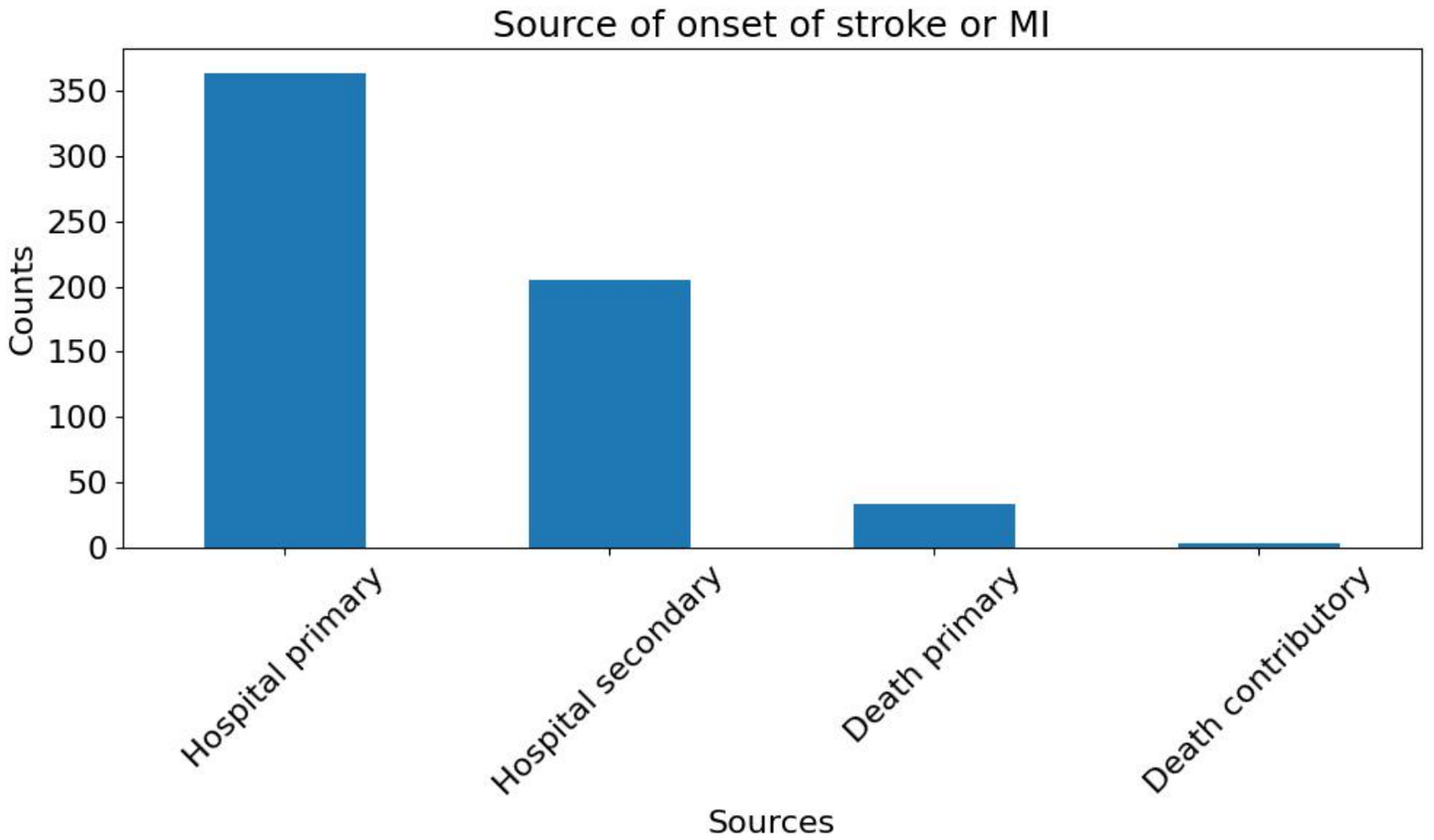}
    \caption{Bar plots showing the methods used to determine the onset date of stroke or MI for the 612 CVD+ subjects involved in our classification task (see details in \citep{ukbiobank_alg_outcomes})}
    \label{fig:source_cvd}
\end{figure}

Figure \ref{fig:source_cvd} presents a bar plot illustrating the methods used to determine the onset date of stroke or myocardial infarction (MI) for the 612 \texttt{CVD+} subjects involved in our classification task. The majority of cases, approximately 350, were attributed to hospital primary records, followed by hospital secondary records. The fewest cases were associated with death contributory records.

\begin{figure}[H]
    \centering
    \begin{subfigure}[b]{0.45\textwidth}
        \includegraphics[width=\linewidth]{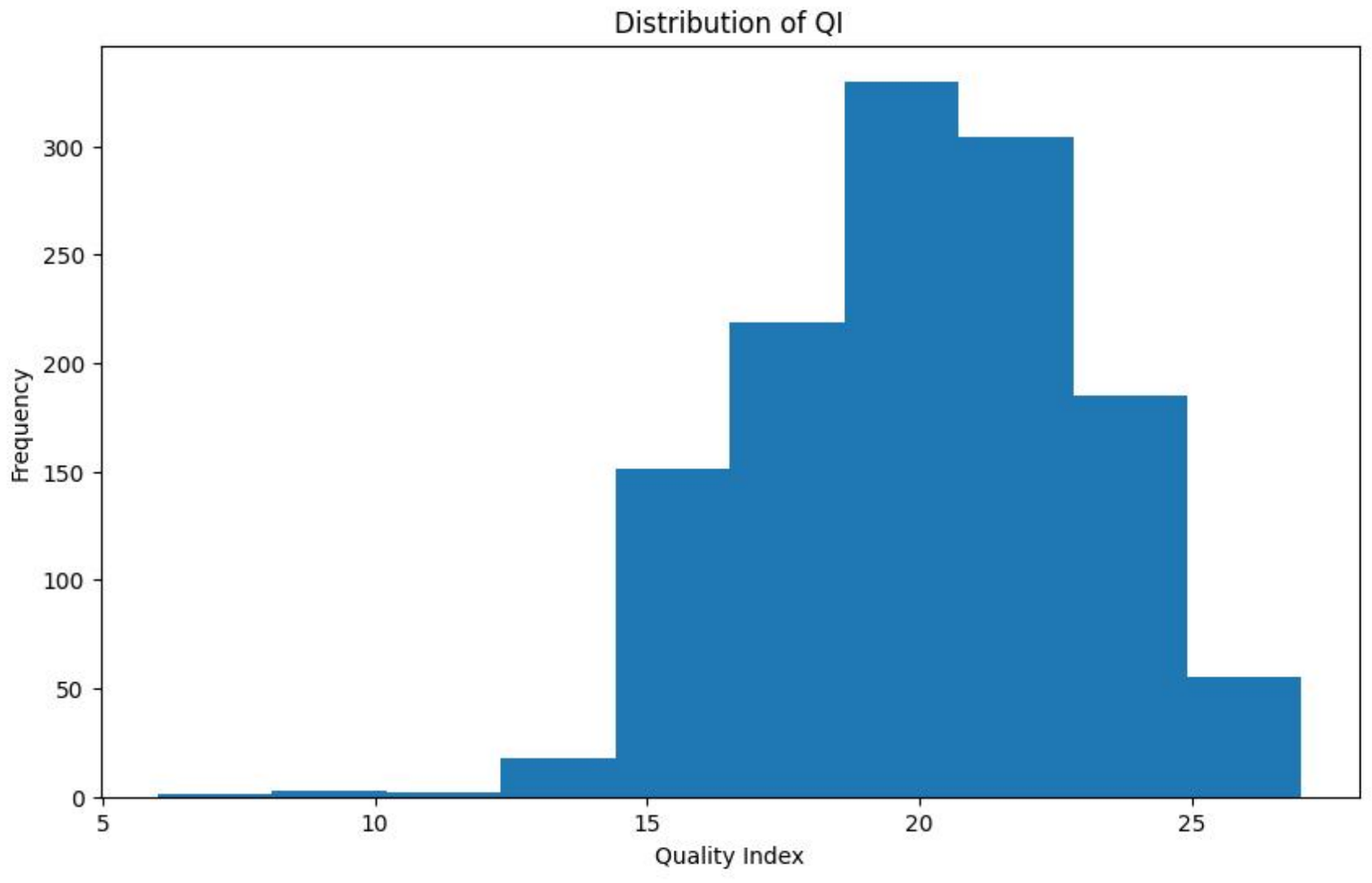}
        \caption{Distribution of QI of the left eye}
        \label{fig:fig1}
    \end{subfigure}
    \hfill
    \begin{subfigure}[b]{0.45\textwidth}
        \includegraphics[width=\linewidth]{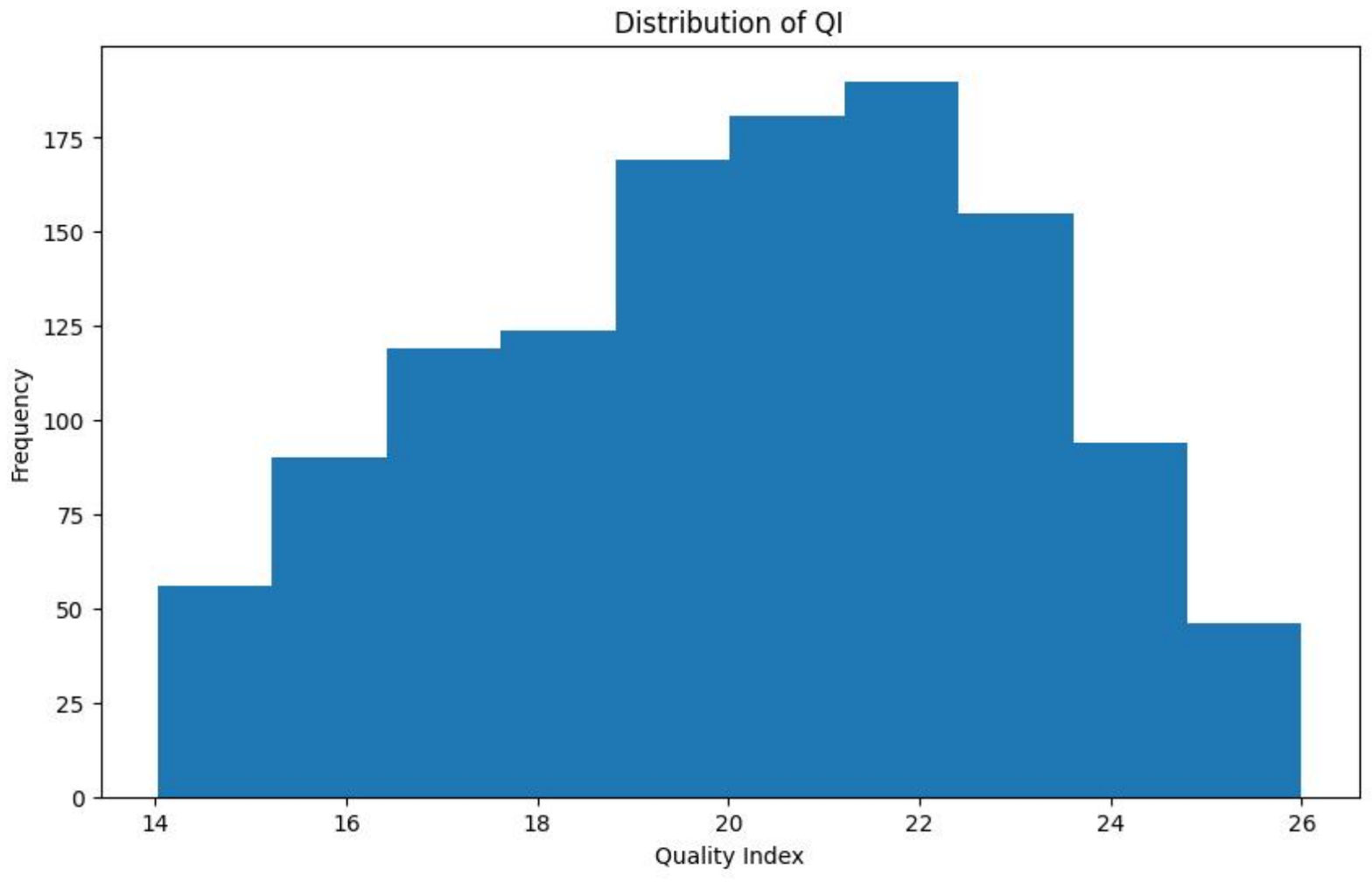}
        \caption{Distribution of QI of the right eye}
        \label{fig:fig2}
    \end{subfigure}
    \caption{Comparison of the distribution of the Quality Index from both eyes.}
    \label{fig:qis}
\end{figure}

The Quality Index is calculated as the product of two terms referred to as  Intensity Ratio (IR) and Tissue Signal Ratio (TSR). The IR is akin to the signal-to-noise ratio (SNR), but rather than considering the maximum SNR value among all A-scans, it encompasses the entire image. Meanwhile, TSR represents the ratio of highly reflective pixels to those with lower reflectivity. Further details regarding the formula are provided in \citep{Stein2006ANQ}. The QI for both left and right eye OCT imaging is shown in Figure \ref{fig:qis}.

\begin{figure}[H]
    \centering
    \includegraphics[width=0.91\textwidth]{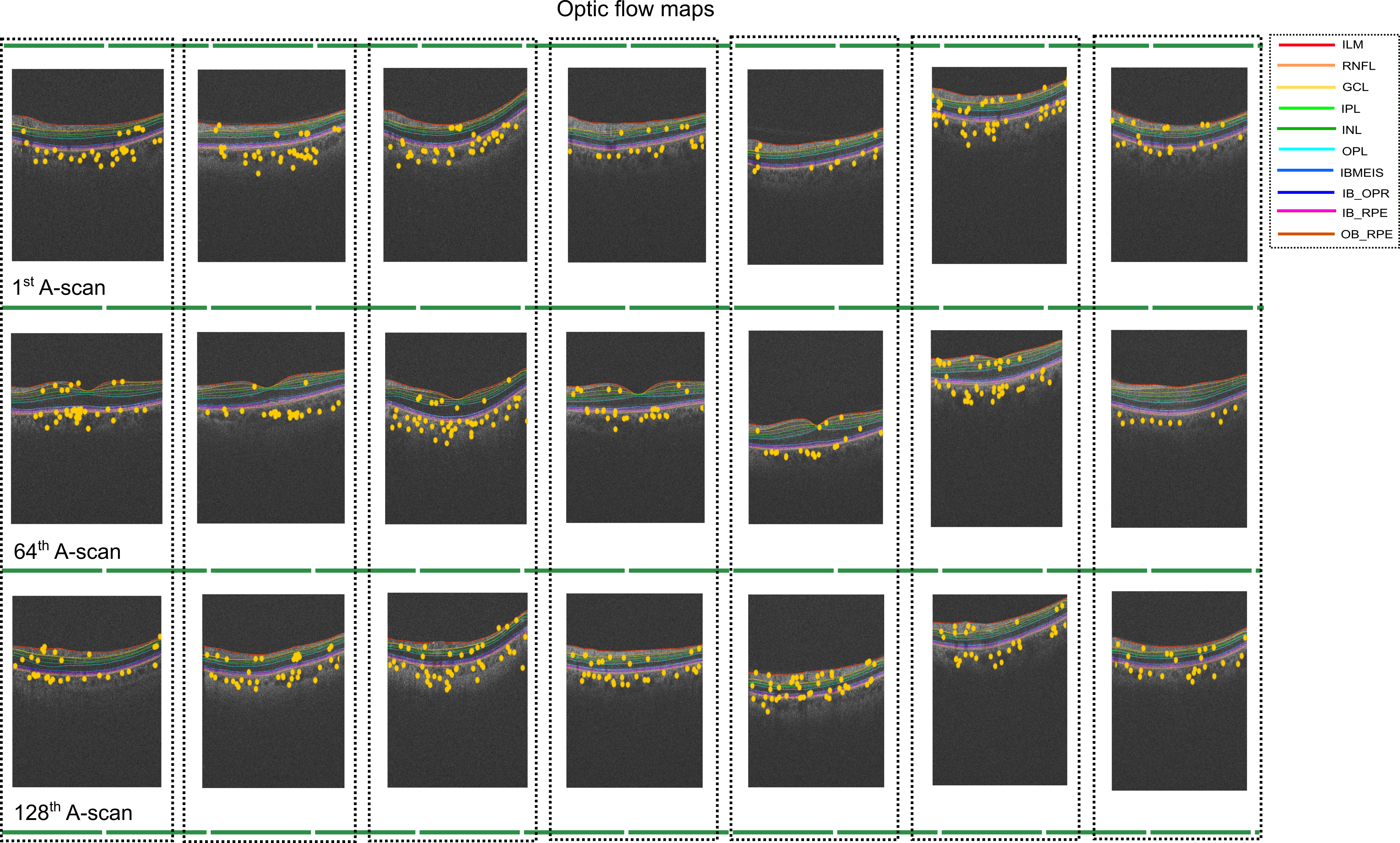}
    \caption{Additional instances of optical flow maps for seven subjects, each displaying three distinct B-scans for the left eye. The top row corresponds to the 1st B-scan, the middle image to the 64th B-scan, and the bottom row depicts the final B-scan. The yellow circles highlight regions of the optical flow maps when modifying the latent variable zl066. The legend includes the names of layer boundaries in the optical coherence tomography images: Internal Limiting Membrane, Retinal Nerve Fiber Layer (RNFL), Ganglion Cell Layer (GCL), Inner Plexiform Layer (IPL), Inner Nuclear Layer (INL), Outer Plexiform Layer (OPL), Boundary of Myoid and Ellipsoid of Inner Segments (BMEIS), Inner Boundary of Outer Segment Retinal Pigment Epithelium Complex (IB \textunderscore OPR), Inner Boundary of Retinal Pigment Epithelium (IB \textunderscore RPE), and Outer Boundary of Retinal Pigment Epithelium (OB \textunderscore RPE).}
    \label{fig:optical_flow_appendix}
\end{figure}

Figure \ref{fig:optical_flow_appendix} presents additional examples of optical flow maps for seven participants.

Table \ref{t:qrisk_variables} presents the list of variables used to calculate QRISK3, as obtained from \citep{Li2019RP}.

\begin{table}[H]
  \centering
  \renewcommand{\arraystretch}{1.4}
  \begin{tabular}{ |p{5cm}|}
    \hline
    \textbf{Variables used in QRISK3}  \\
    \hline
     Sex\\
     \hline
     Age\\
     \hline
     Atrial fibrillation\\
     \hline
     Atypical antipsy\\
     \hline
     Regular steroid tablets\\
     \hline
     Erectile disfunction\\
     \hline
     Migraine\\
     \hline
     Rheumatoid arthritis\\
     \hline
     Chronic kidney disease\\
     \hline
     Severe mental illness\\
     \hline
     Systemic lupus erythematosis\\
     \hline
     Blood pressure treatment \\
     \hline
     Diabetes1 \\
     \hline
     Diabetes2 \\
     \hline
     Weight \\
     \hline
     Height \\
     \hline
     Ethnicity \\
     \hline
     Heart attack relative \\
     \hline
     Cholesterol HDL ratio \\
     \hline
     Systolic blood pressure \\
     \hline
     Std systolic blood pressure \\
     \hline
     Smoke \\
     \hline
     Townsend \\
     \hline
  \end{tabular}
  \caption{List of Variables used to calculate QRISK3}
  \label{t:qrisk_variables}
\end{table}

Table \ref{t:chisquared_all} presents the chi-squared ($\chi^2$) test results comparing BE-MTDT-RF with QRISK, LE-MTDT-RF, RE-MTDT-RF, MTDT-RF, BE-RF, LE-RF, and RE-RF, as well as BE-RF with LE-RF and RE-RF. The $p$-values indicate statistically significant differences between these classifiers (*$p < 0.005$). Table \ref{t:chisquared_left_right} presents the chi-squared ($\chi^2$) test results comparing LE-MTDT-RF and RE-MTDT-RF ($9.61 \times 10^{-5}$), and LE-RF and RE-RF ($4.22 \times 10^{-6}$), with $p-values < 0.005$.

\begin{table}[H]
  \centering
  \renewcommand{\arraystretch}{1.4}
  \begin{tabular}{ |p{3cm}|p{3cm}|p{2cm}|p{3cm}|}
    \hline
    Classifier 1 & Classifier 2 & $\chi^2$ & P value \\
    \hline
    BE-MTDT-RF & QRISK & $95.72$ & $1.32 \times 10^{-22}$ \\
    BE-MTDT-RF & LE-MTDT-RF & $15.21$ & $9.261 \times 10^{-05}$ \\
    BE-MTDT-RF & RE-MTDT-RF & $21.16$ & $4.22\times 10^{-06}$ \\
    BE-MTDT-RF & MTDT-RF & $93.23$ & $4.65 \times 10^{-22}$ \\
    BE-MTDT-RF & BE-RF & $10.59$ & $0.0011$ \\
    BE-MTDT-RF & LE-RF & $27.43$ & $1.63 \times 10^{-07}$ \\
    BE-MTDT-RF & RE-RF & $43.83$ & $3.57 \times 10^{-11}$ \\
    BE-RF & LE-RF & $21.16$ & $4.22 \times 10^{-06}$ \\
    BE-RF & RE-RF & $37.78$ & $7.89 \times 10^{-10}$ \\
    \hline
  \end{tabular}
  \caption{Comparison between various classifiers based on different configurations of eye data (both eyes, left eye, right eye) and metadata inclusion. The chi-squared ($\chi^2$) values and associated p-values indicate the statistical significance of differences between these classifiers.}
  \label{t:chisquared_all}
\end{table}

\begin{table}[H]
  \centering
  \renewcommand{\arraystretch}{1.4}
  \begin{tabular}{ |p{3cm}|p{3cm}|p{2cm}|p{3cm}|}
    \hline
    Classifier 1 & Classifier 2 & $\chi^2$ & P value \\
    \hline
    LE-MTDT-RF & RE-MTDT-RF & $6$ & $9.61 \times 10^{-05}$ \\
    LE-RF & RE-RF & $17.06$ & $4.22 \times 10^{-06}$ \\
    \hline
  \end{tabular}
  \caption{Comparison of LE-MTDT-RF and RE-MTDT-RF, as well as LE-RF and RE-RF, using McNemar's Test (*$p < 0.005$).}
  \label{t:chisquared_left_right}
\end{table}

Supplementary Table \ref{t:metrics_mlp} provides the classification performance metrics evaluated on the same holdout set used for the experiments reported in Table \ref{t:metrics}. In this comparison, we assessed the performance of the BE-MTDT-RF model, which achieved superior results, against a Multilayer Perceptron (MLP) algorithm. For the MLP model, we conducted a comprehensive grid search to identify its optimal configuration, ensuring a fair comparison. Both models were evaluated using ocular data and metadata as input features. The BE-MTDT-RF model demonstrated superior performance across all evaluated metrics. We selected Random Forest (RF) as our classifier model due to its ability to mitigate overfitting \citep{Breiman2001RandomF}. Additionally, the feature importance analysis provided by RF offers valuable insights into the relative contributions of different features to the model's predictions, aligning with our exploratory objectives and potentially enhancing domain-specific understanding of the dataset. Moreover, RF is computationally less demanding than neural networks and does not require GPU resources for training, making it a more efficient choice for our analysis.

\begin{table}[H]
  \centering
  \renewcommand{\arraystretch}{1.4}
  \begin{tabular}{ |p{2.5cm}|p{2cm}|p{2cm}|p{2cm}|p{2cm}|}
    \hline
    \multicolumn{5}{|c|}{Classification metrics} \\
    \hline
    Modality & Accuracy & Sensitivity & Specificity & AUC-Value \\
    \hline
    \textbf{BE-MTDT-RF} & \textbf{0.7} & \textbf{0.7} & \textbf{0.7} & \textbf{0.75} \\
       MLP & 0.65 & 0.635 & 0.65 & 0.719\\
    \hline
  \end{tabular}
  \caption{Comparison of classification metric results between our model employing both ocular data and metadata (BE-MTDT-RF) and the MLP algorithm.}
  \label{t:metrics_mlp}
\end{table}

Table \ref{t:vae_architectures} shows the architectural design of the Variational Autoencoder (VAE) used in our model, detailing the encoder and decoder components. It presents the structure of each network, including the layer types, configurations, and dimensional transformations that occur between layers. The encoder extracts features from the input RGB image, while the decoder reconstructs the image from the latent variables.

\begin{table}[H]
\centering
\renewcommand{\arraystretch}{1.5} 
\begin{tabular}{ |p{6cm}|p{6cm}| }
    \hline
    \textbf{Encoder} & \textbf{Decoder} \\
    \hline\hline
    Input: $128 \times 224 \times 224$ (RGB image) & Latent variables, Fully Connected, Reshape \\
    \hline
    Conv2D(128, $3 \times 3$, stride=$2 \times 2$, padding=$1 \times 1$) & Conv2D(64, $3 \times 3$, stride=$2 \times 2$, padding=$1 \times 1$) \\
    BatchNorm2d(128) & BatchNorm2d(64) \\
    ReLU & ReLU \\
    \hline
    Conv2D(256, $3 \times 3$, stride=$2 \times 2$, padding=$1 \times 1$) & Conv2D(64, $3 \times 3$, stride=$2 \times 2$, padding=$1 \times 1$) \\
    BatchNorm2d(256) & BatchNorm2d(64) \\
    ReLU & ReLU \\
    \hline
    Conv2D(128, $3 \times 3$, stride=$2 \times 2$, padding=$1 \times 1$) & Conv2D(128, $3 \times 3$, stride=$2 \times 2$, padding=$1 \times 1$) \\
    BatchNorm2d(128) & BatchNorm2d(128) \\
    ReLU & ReLU \\
    \hline
    Conv2D(128, $3 \times 3$, stride=$2 \times 2$, padding=$1 \times 1$) & Conv2D(128, $3 \times 3$, stride=$2 \times 2$, padding=$1 \times 1$) \\
    BatchNorm2d(128) & BatchNorm2d(128) \\
    ReLU & ReLU \\
    \hline
    Conv2D(64, $3 \times 3$, stride=$2 \times 2$, padding=$1 \times 1$) & Conv2D(256, $3 \times 3$, stride=$2 \times 2$, padding=$1 \times 1$) \\
    BatchNorm2d(64) & BatchNorm2d(256) \\
    ReLU & ReLU \\
    \hline
    Conv2D(64, $3 \times 3$, stride=$2 \times 2$, padding=$1 \times 1$) & Conv2D(128, $3 \times 3$, stride=$2 \times 2$, padding=$1 \times 1$) \\
    BatchNorm2d(64) & BatchNorm2d(128) \\
    ReLU & ReLU \\
    \hline
    Latent variables, Fully Connected & Output: $128 \times 224 \times 224$ (Reconstructed RGB image) \\
    \hline
\end{tabular}
\caption{Architectural design of the Variational Autoencoder (VAE), showing the encoder and decoder structures with their respective layers, configurations, and dimensional transformations.}
\label{t:vae_architectures}
\end{table}

The Excel file titled \texttt{Supplementary\_Grid\_Search\_Results\_VAE\_RF\_MLP.xlsx} contains the results from the grid search experiments conducted to determine the optimal parameters for our models (VAE, Random Forest, and Multilayer Perceptron). The file includes multiple sheets corresponding to different experiments. For example, the sheet \texttt{oct\_vae\_433} presents the grid search results for the VAE model with a data split ratio of 4:3:3 (training/validation/testing). Similarly, the sheet \texttt{oct\_bemtdt\_rf\_433} provides the parameters and results from the grid search using the \texttt{sklearn.RandomizedSearchCV} library for the Random Forest model. Additionally, the file contains a comparison of the best results across various data split ratios, allowing for a clearer understanding of the performance differences.

The code used in this study is available on GitHub at the following address: \href{https://github.com/CMaldonado1/Predicting-CVD-using-OCT-images}{Github/Predicting-CVD-using-OCT-images}


\end{document}